\documentclass[11pt]{article} % use larger type; default would be 10pt
\pdfoutput=1 % if your are submitting a pdflatex (i.e. if you have
\usepackage{jheppub} % for details on the use of the package, please
\usepackage[utf8]{inputenc}
\usepackage[english]{babel}
\usepackage{amsmath,amssymb,amsbsy,amstext,amsthm,booktabs,simplewick,exscale,relsize,slashed,graphicx,amsfonts,upgreek, xcolor,subcaption}
\usepackage{multirow,color,dcolumn,bm,enumerate}
\usepackage{hyperref}
\usepackage{feynmp-auto,axodraw2,tikz}
\usepackage{ulem}
\usepackage{comment}

\usepackage{hyperref}
\usepackage[capitalise,noabbrev,nameinlink]{cleveref}

\title{Counting Operators in $N=1$ Supersymmetric Gauge Theories}
\author[a]{Antonio Delgado,}
\author[a]{Adam Martin,}
\author[a]{Runqing Wang}

\affiliation[a]{Department of Physics, University of Notre Dame,
  South Bend, IN, 46556 USA}

\emailAdd{adelgad2@nd.edu}
\emailAdd{amarti41@nd.edu}
\emailAdd{rwang7@nd.edu}

\abstract{Following a recent publication, in this paper we count the number of independent operators at arbitrary mass dimension in $N=1$ supersymmetric gauge theories and derive their field and derivative content. This work uses Hilbert series machinery and extends a technique from our  previous work on handling integration by parts redundancies to vector superfields. The method proposed here can be applied to both abelian and non-abelian gauge theories and for any set of (chiral/antichiral) matter fields. We work through detailed steps for the abelian case with single flavor chiral superfield at mass dimension eight, and provide other examples in the appendices.
}

\begin{document}
\maketitle

\setcounter{page}{2}

%\tableofcontents 
\section{Introduction}
\label{intro}
Higher dimensional operators are a key ingredient in effective field theories (EFT), they encode our lack of knowledge on the theory above a scale $\Lambda_E$ \footnote{ Not to be confused with a chiral superfield $\Lambda$ that appears in supersymmetric gauge transformations.}. These non-renomalizable operators are  built out of combinations of field and derivatives. However not all operators you can build for a given dimension are  linearly independent under the equivalence relations defined by integration by parts (IBP) and equations of motion (EOM). It is therefore important to be able to give a complete list of {\it independent} operators once those redundancies are taken into account.  

It has  been known for a some time  that in the non-supersymmetric case, one can remove both IBP and EOM redundancies using a Hilbert series approach \cite{Lehman:2015via, Henning:2015daa,Lehman:2015coa,Henning:2015alf, Henning:2017fpj,Anisha:2019nzx,Marinissen:2020jmb, Graf:2020yxt,Wang:2021wdq,Graf2023,Yu2022,Yu2022a,Yu2021,Lu2021,Lu2022,Sun2022,Trautner_2019,Trautner_2020}, originating from the deep connection between conformal group/algebra and operator basis \cite{Henning:2015daa, Henning:2015alf, Henning:2017fpj}. Several other methods, such as the on-shell Young Tableau construction~\cite{Henning:2019enq, Henning:2019mcv, Ma:2019gtx, Durieux:2019siw, Dong:2021yak,Li:2022tec,Fonseca:2019yya,Li:2020gnx}, also serve as complementary ways to give the number and explicit form of these operators. However, neither of these methods can be directly generalized to the supersymmetric theory, due to the existence of 4 additional Grassmannian variables (for $\mathcal{N}=1$, 4D supersymmetry), which induces 4 extra superderivatives. As a result, the usual way to remove IBP and EOM relations no longer works in this scenario. 

Nevertheless, an approach based on Hilbert series techniques was recently put forward  \cite{Delgado:2022bho} to count the  number of independent operators in supersymmetric theories. It showed how the existence of two superderivatives implies IBP-like relations among operators that do not appear in non-supersymmetric theories. The method allows one to find the operator basis for arbitrary mass dimensions. In Ref.~\cite{Delgado:2022bho}, the method was demonstrated for theories with chiral and anti-chiral superfields only, hence the goal of this paper is to extend the formalism to include vector superfields and gauge interactions.

Supersymmetric gauge theories with higher dimensional operators as EFTs have been studied before \cite{Antoniadis:2007xc,Forcella:2008tqe,Farakos:2013zsa,Dudas:2015vka, Nitta:2020gam}, including supersymmetry breaking, equivalence between certain higher dimensional operators with derivatives and second order supersymmetry (without derivatives) with additional superfields, etc. However,  a complete analysis of the full operator basis hasn't been studied yet. In this paper we calculate the number of different independent operators  by studying abelian supersymmetric gauge theories first, and then explore the non-abelian case using the previously built Hilbert series method.

 In non-supersymmetric theories, adding gauge/internal symmetry groups into the theory is accommodated by adding additional Haar measures, characters, and related group parameters (complex, unit modulus numbers parametrizing elements of the maximal torus of the group) into the corresponding Hilbert series. However, as we will see in this paper, in a supersymmetric gauge theory we need to double the group parameters in order to make use of the Hilbert series. This doubling originates from the fact that the gauge transformations in supersymmetry are defined through superfields instead of real-number phases, and is needed to ensure that the Hilbert series do not generate unwanted terms.

The paper is organized as follows: In Section \ref{results} we will give a short introduction of the approach of Ref. \cite{Delgado:2022bho}, explaining the meaning of each term in the master formula Eq.~\eqref{master}. Then we apply it in an abelian supersymmetric gauge theory, defined in Section \ref{abelian}, where we work out an example explicitly. In Section \ref{non abelian} we make a generalization to the non-abelian case and we work through an example as well. Section \ref{conclusion} provides a short discussion and possible future applications. We will follow the notations and conventions the same as our previous paper \cite{Delgado:2022bho}\footnote{Through out the paper we adopt the most-negative metric tensor in Minkowski space, i.e. $\eta^{\mu\nu}=\eta_{\mu\nu}=diag(+1, -1, -1, -1)$. Totally antisymmetric tensor in two dimensions $\epsilon_{AB}(A,B=1,2)$ are defined to be $\epsilon_{12}=\epsilon^{21}=1; \epsilon_{21}=\epsilon^{12}=-1$. In addition, a useful identity we will use is $\epsilon_{AB}\epsilon_{CD}+\epsilon_{AC}\epsilon_{DB}+\epsilon_{AD}\epsilon_{BC}=0$.}.

\section{Methods and Results}\label{results}

\subsection{Methods}\label{methods}

Hilbert series are a useful tool for determining the number of group invariants~\cite{Pouliot:1998yv, Benvenuti:2006qr,Butti:2007jv,Feng:2007ur,Forcella:2007wk, Dolan:2007rq,Gray:2008yu,Hanany:2008sb,Benvenuti:2010pq,Chen:2011wn,Hanany:2012dm,Rodriguez-Gomez:2013dpa,Dey:2013fea,Hanany:2014hia,Hanany:2014dia}. Formally, Hilbert series are a power series $\mathcal H = \sum\limits_n c_n t^n$ where $c_n$ is the number of invariants at order $t^n$. Applied to field theory (both supersymmetric and non-supersymmetric), the invariants are operators that are singlets under all Lorentz and internal symmetries, and the order of the invariant is set by the mass dimension of the operator. By construction, the Hilbert series accounts for group redundancies (e.g. operator relations via Fierz transformations) when combined with Haar measure. However, when counting operators with derivatives, additional ingredients are required to remove operator redundancies due to equations of motion (EOM) and integration by parts (IBP). The IBP relations are where the main differences between supersymmetric and non-supersymmetric theories lie, and were the topic of Ref.~\cite{Delgado:2022bho}. Adjusted to remove these redundancies related to derivatives, the master Hilbert series formula for counting invariants in a supersymmetric theory of chiral and antichiral fields is:
\begin{equation}\label{master}
\mathcal{H}(P,Q,\{\Phi_a\})=\int d\mu_{Lorentz}d\mu_{gauge}d\mu_{U_R(1)}P^{-1}(P,Q,\alpha,\beta,z)PE[\sum_a\Phi_a\tilde{\chi}_a\chi'_a].
\end{equation}
Here, $\{\Phi_a\}$ are a set of spurions representing the chiral/antichiral superfields in the theory, i.e. $\{\Phi_a\}=\{\Phi_1,\Phi_2,\cdots\Phi_1^\dagger,\Phi_2^\dagger,\cdots\}$. The fields, along with their Lorentz and internal symmetry characters $\tilde \chi_a, \chi'_a$, respectively, are grouped into a function -- the plethystic exponential (PE) -- that generates all possible products of $\{\Phi_a\}$ and their derivatives. The other ingredients in Eq.~\eqref{master} are $d\mu_G$, the Haar measure of group $G$, and $P^{-1}(P,Q,\alpha,\beta,z)$ a factor that removes IBP redundancies. The arguments of the IBP factor include spurions $P, Q$ representing the superderivatives $D_\alpha,\overline{D}_{\dot{\alpha}}$, and their group characters.

The output of Eq.~\eqref{master} is a graded Hilbert series, meaning it returns not only the number of invariant operators at each mass dimension, but also their field content -- the number of $\Phi_i, \Phi^\dag_i, D_\alpha, \bar D_{\dot \alpha}$. While the field content of an invariant is known, no information on exact placement of the derivatives within the operator or how all indices are contracted is provided by the Hilbert series and must be worked out by other means.

Having shown the master formula, let us now give a few more details on each of its main ingredients. \\

$\bullet$ \textbf{\textit{Plethystic Exponential and Haar Measure}} \\

The plethystic exponential generates all products of its arguments consistent with the statistics (symmetric or antisymmetric) of each ingredient. For a single argument, it is defined as 
\begin{equation}
PE[\Phi_R ]=\exp\Big\{\sum_n\frac{1}{n}(\pm1)^{n+1}\Phi_R^n\chi^n_{{}_{\mathcal G, R}} \Big\},
\end{equation}
where we choose $+1$ for symmetric products (bosonic fields) and $-1$ for antisymmetric products (fermionic fields). Each argument $\Phi_R$ (for our purposes, a spurion representing a field in the theory) is accompanied by the character $\chi$ corresponding to it's representation ($R$) under all groups ($\mathcal G$) of the theory\footnote{$\chi^n_{\mathcal G, R} $ is a shorthand for $\chi_{\mathcal G, R}(\alpha^n, \beta^n, \cdots)$ where $\alpha, \beta$, etc. are the group parameters for $\mathcal G$.}. The characters encapsulate the transformation properties of each product of $\Phi$ in the PE, e.g. a symmetric product of two $SU(2)$ triplets would be accompanied by the character for the $SU(2)$ triplet representation. For multiple spurions (fields), the net PE is the product of each individual spurion's PE.

To project out products of $\Phi_i$ that transform under certain representation, we can use character orthonormality:
\begin{align}
\int d\mu_{\mathcal G}\, \chi^*_{\mathcal G, I}\, \chi_{\mathcal G,J} = \delta_{IJ},
\end{align}
where $d\mu_{\mathcal G}$ is the Haar measure for the group $\mathcal G$, provided $\mathcal G$ is compact. Specifically, the products of $\Phi_i$ we are most interested in are the $\mathcal G$ invariants, as these form the Hilbert series. These can be obtained by multiplying the PE by the character for the trivial/singlet representation, $\chi_{\mathcal G, trivial} = 1$ and integrating over $d\mu_\mathcal G$. \\

$\bullet$ \textbf{\textit{Conformal Characters: Removing EOM}} \\

Forgetting any internal symmetries for the moment, each field $\Phi_i$ sits in some representation of the Lorentz group, and must carry the character for that representation with it in the PE\footnote{For supersymmetric theories, each $\Phi_i$ also has an R-charge as well.}. Moreover, as explained in Ref.~\cite{Lehman:2015via}, we want to lump all powers of derivatives acting on the field/spurion (minus those removed by EOM) in with each field in the PE. For example, if $\Phi_i$ is a Lorentz scalar, we want to extend the argument of the PE to $\Phi\, \chi_{(0,0)} + (PQ)\partial_\mu \Phi\, \chi_{(\frac 1 2, \frac 1 2)} + (PQ)^2\,\partial_{\{\mu, \nu\}}\Phi \,\chi_{(1,1)} + \cdots$; note that the character changes as we include more derivatives, and we include the spurion product $(PQ)$ to count the number of (super) derivatives\footnote{In a non-supersymmetric theory, a single spurion, $D$ is sufficient to count derivatives.}. The sum over all possible derivatives on $\Phi$ can be carried out, and the results take the form of characters for short representations of the {\it conformal} symmetry group. Viewed from the conformal perspective, each field sits along with all its derivatives in a single infinite-dimensional representation where the initial field is denoted as the primary component and all derivatives as descendants. The conformal characters for several important representations are listed below, denoted as $\tilde \chi_{(j_1, j_2)}$, where $j_1, j_2$ indicated the Lorentz representation of the primary:
\begin{align}
\bar{\chi}_{(0,0)} &= C(\alpha, \beta,P,Q)(1-(PQ)^2)  \label{eq:scal} \\ 
\bar{\chi}_{(\frac 1 2,0)} &= C(\alpha,\beta,P,Q )((\alpha+\frac{1}{\alpha})-(PQ)(\beta+\frac{1}{\beta}))  \label{eq:ferm}  \\
\bar{\chi}_{(0,\frac{1}{2})} &=C(\alpha,\beta,P,Q)((\beta+\frac{1}{\beta})-(PQ)(\alpha+\frac{1}{\alpha}))  \\
\bar{\chi}_{(1,0)} &= C(\alpha,\beta,P,Q )((\alpha^2+1+\frac{1}{\alpha^2})-(PQ)(\alpha+\frac{1}{\alpha})(\beta+\frac{1}{\beta})+(PQ)^2) \label{eq:spin1} \\
\bar{\chi}_{(0,1)} &= C(\alpha,\beta,P,Q )((\beta^2+1+\frac{1}{\beta^2})-(PQ)(\alpha+\frac{1}{\alpha})(\beta+\frac{1}{\beta})+(PQ)^2).
\label{eq:confchar}
\end{align}
The factor $C(\alpha, \beta,P,Q)$ is defined as:
 \begin{align}
C(\alpha, \beta,P,Q) = \Big( (1-PQ\alpha\beta)(1-\frac{PQ}{\alpha\beta})(1-\frac{PQ\alpha}{\beta})(1-\frac{PQ\beta}{\alpha}) \Big)^{-1},
\end{align}
where $\alpha$ and $\beta$ are group parameters related to $SU(2)_L$ and $SU(2)_R$ and $P,Q$ are spurions to represent $D_{\alpha}$ and $\overline{D}_{\dot{\alpha}}$.

The conformal symmetry link was crucial to the understanding of Hilbert series in non-supersymmetric theories~\cite{Henning:2015daa, Henning:2017fpj}. The obvious extension for supersymmetric theories would be to organize things using superconformal representations. However, so far we have not found this direction to be fruitful (see Ref.~\cite{Delgado:2022bho} for some discussion on this point). Therefore, for now we will utilize the $\tilde \chi$ simply as neat packages of fields and their derivatives, forgetting their connection to conformal field theory.

A further complication for the derivative ``tower" for superfields is the application of a superderivative $D_\alpha, \bar D_{\dot{\alpha}}$ to a superfield flips the statistics of the field, e.g. for a bosonic superfield $\Phi$, $D_\alpha \Phi$ is fermionic. This pattern continues with subsequent derivatives, so that when we want to combine a superfield and all its superderivatives into a term in the PE, we need two terms -- one for the bosonic combinations of field plus superderivatives, and one for the fermionic combinations. Thus, for a bosonic superfield $\Phi$, we add $\Phi \tilde\chi_{(0,0)}$ to the bosonic PE (even numbers of $D_\alpha, \bar D_{\dot{\alpha}}$) and $P(D\Phi)\,\tilde\chi_{(\frac 1 2,0)}$ (odd numbers of $D_\alpha, \bar D_{\dot{\alpha}}$) to the fermionic PE.  Here, we treat $(D\Phi) $ as a different spurion, ignoring any connection to $\Phi$, using the spurion $P$ to track the extra derivative factor contained in $(D\Phi)$.\footnote{For $\Phi^\dag$, the replacement is $\Phi^{\dag} \tilde\chi_{(0,0)}$ in the bosonic PE and $Q(\overline D \Phi^{\dag}) \tilde\chi_{(0,\frac 1 2)}$ in the fermionic PE, where $\bar D\Phi^{\dag}$ is a separate spurion.} The fact that each superfield leads to a term in both the bosonic and fermionic PE is a manifestation of supersymmetry.  \\

$\bullet$ \textbf{\textit{$\boldsymbol{P^{-1}(p,q,\alpha,\beta,z)}$ Factor: Removing IBP}} \\

The last piece needs to be clarified in \eqref{master} is the $P^{-1}(P,Q,\alpha,\beta,z)$ factor, which is explicitly given by an infinite sum of spurions $P,Q$ as well as group parameters $x=\alpha+\frac{1}{\alpha},y=\beta+\frac{1}{\beta}$:
\begin{equation}\label{P}
\begin{split}
P^{-1}(p,q,\alpha,\beta,z)&=1\\
&-(Px+Qy)\\
&+(PQ^2x+P^2Qy+P^2(x^2-1)+Q^2(y^2-1))\\
&-(PQ^3xy+P^3Qxy+P^3(x^3-2x)+Q^3(y^3-2y)+P^2Q^2)\\
&\cdots
\end{split}
\end{equation}
This factor allows one to fully remove IBP redundancies in $N=1$ supersymmetry, as demonstrated in our previous work \cite{Delgado:2022bho}. The terms in the expansion of \eqref{P} count the relations in `correction spaces' for any given term in the PE.\\

Putting the pieces together, we can now read off what \eqref{master} does. To count the independent operators, one starts by determining the representation (both spacetime and internal) of each superfield, then taking the plethystic exponential of the product of corresponding group characters with a spurion for the field. This allows one to work with the building blocks free of EOM. Adding the $P^{-1}(P,Q,\alpha,\beta,z)$ factor will remove all IBP redundancies, and finally integrating over Haar measure of both spacetime and internal groups projects out the invariant operators. Although the formal expansion of $P^{-1}(P,Q,\alpha,\beta,z)$ \eqref{P} is infinite, it actually terminates at given order due to the fact that one cannot build arbitrary larger representation with finite number of superfields and derivatives. To speed up calculations, it is often beneficial to determine the maximum representation (for a given field/derivative content) before plugging into \eqref{master} and truncating $P^{-1}(P,Q,\alpha,\beta,z)$ appropriately.

There is a subtlety we should mention before moving forward. The conformal characters for the representations listed earlier are not orthonormal, a consequence of the fact the that conformal group is non-compact. As a result, Eq. \eqref{master} contains an unwanted $\Delta H$ piece, which is common both in non-supersymmetric case \cite{Henning:2017fpj} and supersymmetric case \cite{Delgado:2022bho}. However, as proved/argued in these papers, this term only contains operators with mass dimensions less than or equal to four, and is therefore irrelevant if our goal is to determine the operator basis for higher dimensional operators (dimension $\geq5$). As a result, we will ignore this term in the rest of this paper.

Having reviewed \eqref{master}, we are now prepared to include gauge interactions, where one needs to consider gauge invariance in addition to Lorentz symmetry and R-symmetry. To get familiar with how the procedure works, we first discuss the abelian case in the next section, and then move to non-abelian case in section \ref{non abelian}. 

\subsection{Abelian supersymmetric gauge theory}\label{abelian}

In an $N=1$ supersymmetric $U(1)$ gauge theory, chiral superfields $\Phi_l$ transform as,
\begin{equation}\label{u1trans}
\Phi_l\rightarrow\Phi'_l=e^{-it_l\Lambda}\Phi_l;\ \Phi_l^{\dagger}\rightarrow\Phi_l^{'\dagger}=e^{it_l{\Lambda^\dagger}}\Phi^{\dagger}_l,
\end{equation} 
where  $t_l$ is a real number (identified as the gauge charge) and $\Lambda, \Lambda^\dagger$ are chiral and antichiral superfields; $\Lambda$ and $\Lambda^\dag$ must be superfields in order for the transformed $\Phi'_l$  ($\Phi_l^{'\dagger}$) to remain chiral (antichiral). To build a gauge invariant term out of these chiral and antichiral superfields, we need to introduce a vector superfield $V$ that transforms as $V\rightarrow V'=V+S+S^\dagger$, where $S$ is a chiral superfield. Setting $S=i\Lambda$ and $S^\dagger=-i\Lambda^\dagger$, we find that the following term (for a single flavor) is gauge invariant under the $U(1)$ local transformations \eqref{u1trans}:
\begin{equation}\label{u1 dterm}
\Phi^\dagger e^{tV}\Phi\rightarrow\Phi^{'\dagger}e^{tV'}\Phi'=\Phi^\dagger e^{tV}\Phi.
\end{equation}
This term can be treated as a generalization of the K\"ahler term $\Phi^\dagger\Phi$ without gauge interactions. To build a gauge invariant term out of the vector superfield alone -- the generalization of the field strength $F^{\mu\nu}$ in a non-supersymmetric case -- one constructs the following:
\begin{equation}\label{W}
W_\alpha\equiv -\frac{1}{4}\overline{D}^2D_\alpha V,\ \ \overline{W}_{\dot{\alpha}}\equiv -\frac{1}{4}D^2\overline{D}_{\dot{\alpha}}V.
\end{equation}
The transformation laws $W_\alpha\rightarrow W'_\alpha=W_\alpha,\overline{W}_{\dot{\alpha}}\rightarrow\overline{W}'_{\dot{\alpha}}=\overline{W}_{\dot{\alpha}}$ follow from the gauge transformation of vector superfield defined above.

The renormalizable Lagrangian which is invariant under this local $U(1)$ symmetry is then formed as: 
\begin{equation}
\mathcal{L}=\frac{1}{4}(W^\alpha W_\alpha+\overline{W}_{\dot{\alpha}}\overline{W}^{\dot{\alpha}})_{\mathcal{F}}+(\Phi_l^\dagger e^{t_lV}\Phi_l)_{\mathcal{D}}+[(\frac{1}{2}m_{ij}\Phi_i\Phi_j+\frac{1}{3}g_{ijk}\Phi_i\Phi_j\Phi_k)_{\mathcal{F}}+h.c.],
\end{equation}
where the subscripts $\mathcal{D,F}$ represent $D$-term and $F$-term by taking the Grassmann integration with $d^4\theta$ and $d^2 \theta$ respectively.

The final object we will need in the superderivative. When there is no gauge symmetry present, recall that the two superderivatives $D_\alpha,\ \overline{D}_{\dot{\alpha}}$:
\begin{subequations}\label{chiral} are defined as:
\begin{align}
&D_{\alpha}=\frac{\partial}{\partial \theta^\alpha}-i\sigma^{\mu}_{\alpha\dot{\alpha}}\overline{\theta}^{\dot{\alpha}}\partial_\mu,\\
&\overline{D}_{\dot{\alpha}}=-\frac{\partial}{\partial\theta^{\dot{\alpha}}}+i\theta^\alpha \sigma^{\mu}_{\alpha\dot{\alpha}}\partial_\mu,
\end{align}
\end{subequations}
where $\theta_\alpha$ and $\overline{\theta}^{\dot{\alpha}}$ are two dimensional Grassmann numbers, and $\partial_\mu$ is the usual partial derivative. They satisfy the following anticommutation relation:
\begin{equation}\label{anti relation}
\{D_{\alpha},\overline{D}_{\dot{\alpha}}\}=2i\sigma^{\mu}_{\alpha\dot{\alpha}}\partial_\mu.
\end{equation}
When acting on superfields transform as \eqref{u1trans}, these derivatives need to be covariantized to respect both supersymmetry and local $U(1)$ symmetry. For this purpose, we define the action of covariant supersymmetric derivatives $\nabla_\alpha$ and $\overline{\nabla}_{\dot{\alpha}}$ on superfields as:
\begin{subequations}\label{derivative}
\begin{align}
&\nabla_\alpha\Phi\equiv(D_\alpha+D_\alpha V)\Phi,\ \nabla_\alpha\Phi^\dagger\equiv D_\alpha\Phi^\dagger=0,\\
&\overline{\nabla}_{\dot{\alpha}}\Phi^\dagger\equiv(\overline{D}_{\dot{\alpha}}+\overline{D}_{\dot{\alpha}} V)\Phi^\dagger,\ \overline{\nabla}_{\dot{\alpha}}\Phi\equiv \overline{D}_{\dot{\alpha}}\Phi=0.
\end{align}
\end{subequations}
Here, $\Phi$ is a chiral superfield and $\Phi^\dagger$ is an antichiral superfield. These two covariant superderivatives also respect supersymmetry, i.e. $\{\nabla_\alpha,\overline{\nabla}_{\dot{\alpha}}\}=2i\sigma^{\mu}_{\alpha\dot{\alpha}}\nabla_\mu$, where $\nabla_\mu$ is the usual Lorentz covariant derivative.  The equation of motion for these super field strengths are given by:
\begin{equation}\label{eom of W}
\nabla^\alpha W_\alpha+\frac{\partial P(V)}{\partial V}=0,
\end{equation}
where $P(V)$ is the term that couples the supercurrent to the vector superfield.

Finally, $N=1$ supersymmetric gauge theories contain a $U(1)_R$ symmetry under which the Grassmann parameter $\theta$ carries charge $+1$ and $\bar{\theta}$ carries charge $-1$. This assignment dictates that $D_\alpha$ has $R$-charge $-1$, $\overline{D}_{\dot{\alpha}}$ has charge $+1$. Unlike chiral/antichiral superfields, which may have any $R$ charge, the $R$ charge of the gauge superfield strengths is set by the kinetic term; $+1$ for $W_\alpha$ and $-1$ for $\overline{W}_{\dot\alpha}$\footnote{When specifying  a superfield's $R$ charge, we are referring the charge of the lowest component field. As $R$ symmetry does not commute with supersymmetry, higher component fields will have higher or lower charge.}.

\subsection{Operator counting for supersymmetric abelian gauge theory}\label{abeliancount}

Adapting the program and main results from Ref.~\cite{Delgado:2022bho} to a $N=1$ supersymmetric abelian gauge symmetry involves a few complications.
\begin{itemize}
\item The main ingredient in the Hilbert series is the plethysm of the relevant degrees of freedom (and derivatives of them), with each spurion accompanied by its representative field's group characters. For the theories studied in Ref.~\cite{Delgado:2022bho}, the basic building blocks were (potentially multiple flavors of) chiral and antichiral fields. For gauge theories, the superfield $V$ appears (and has nice/linear transformation properties) within the field strengths $W_\alpha$ and $\overline{W}_{\dot{\alpha}}$ {\it and} $e^V$. So, do we include all of these, or is there some double-counting given that it is the same $V$ in all three?
\item The second complication is that the gauge transformation parameters in supersymmetric theories are full chiral superfields rather than real numbers. How do we implement this difference in terms of the field/spurion characters, which carry out the group multiplication in the field/spurion plethysm? As an example that emphasizes the role this can play, consider $e^V$, which becomes $e^{-i\Lambda^\dag}e^V e^{i\Lambda}$ under a gauge transformation (setting $t = 1$). If we were to count net $U(1)$ charge, $V$ should have charge zero -- as assignment justified either by what happens in the non-supersymmetric case, or by counting $ \Lambda$ as a real valued constant. However, by this counting, $e^V$ carries no quantum numbers and is dimensionless, so there is nothing to prevent the Hilbert series from adding arbitrary powers of $e^V$ to any operator.
\end{itemize}

Ignoring these subtleties for a moment, let us proceed exactly as in Ref.~\cite{Delgado:2022bho} and identify the relevant degrees of freedom and their Lorentz representations. By looking to the lowest component, we see $\Phi,\Phi^\dagger$ and $e^V\,\sim(0,0)$; $W_\alpha\sim(\frac{1}{2},0);\ \overline{W}_{\dot{\alpha}}\sim(0,\frac{1}{2})$. Next, we identify the representations of the (covariant) derivatives acting on these objects. From our experience with chiral/antichiral theories, we know that odd numbers of acting on a superfield change the statistics of a field, e.g. $\Phi$ is a scalar, while $\nabla_\alpha \Phi$ is a fermion, which must be reflected in how the spurion/field appears in the PE. From our past work, we know $\nabla_\alpha\Phi\sim(\frac{1}{2},0),\ \overline{\nabla}_{\dot{\alpha}}\Phi^\dagger\sim(0,\frac{1}{2})$. Applying the same logic to superfield strengths, $\nabla_{(\beta}W_{\alpha)}\sim(1,0);\ \nabla_{(\dot{\beta}}\overline{W}_{\dot{\alpha})}\sim(0,1)$, where the $(\cdots)$ subscript represents the symmetrization among spinorial indices, e.g. $\nabla_{(\beta}W_{\alpha)}=\nabla_{\beta}W_{\alpha}+\nabla_{\alpha}W_{\beta}$, etc. This leaves derivatives acting on $e^V$, which seems more complicated. However, consider the following:
\begin{align}
\nabla_\alpha(\Phi^\dagger e^{V}\Phi) =D_\alpha(\Phi^\dagger e^{V}\Phi) & \rightarrow \nonumber \Phi^\dag \nabla_\alpha(e^V \Phi) =   \Phi^{\dag}D_\alpha(e^V\Phi) \nonumber
\end{align}
which follows from gauge invariance and the fact that $\Phi^\dag$ is antichiral (Eq.~\eqref{derivative}). Dropping the $\Phi^\dag$, this becomes:
\begin{align}
(\nabla_\alpha V)e^V\Phi + e^V(\nabla_\alpha \Phi) & = (D_\alpha V)e^V\Phi + e^V(D_\alpha \Phi) = e^V(D_\alpha + (D_\alpha V))\Phi \equiv e^V(\nabla_\alpha \Phi), \nonumber 
\end{align}
from which we conclude that $\nabla_\alpha V=0$.  By the same logic, one can show that $\overline{\nabla}_{\dot{\alpha}} V=0$, and therefore that $\nabla_\alpha e^V \equiv \nabla_{\dot \alpha}e^V \equiv 0$. Since all covariant derivatives are zero when acting on $e^V$, we can neglect them from the PE,
\begin{align}
PE[\Phi, \Phi^\dag, \nabla^n \Phi_i, \nabla^n_{\dot\alpha}\Phi^\dag_i, W_\alpha, \overline W_{\dot \alpha}, \nabla^n_\alpha W_\beta, \nabla^n_{\dot \alpha} \overline W_{\dot \beta}, e^V],
\label{eq:superPEarg}
\end{align}
where $i$ labels different matter superfields.

This omission of $e^V$ derivatives clears up one of the subtleties raised at the beginning of the section. The dynamics of the vector superfield $V$ are contained in the superfield strengths and their derivatives, while $e^V$ only plays the role of maintaining gauge invariance. In other words, the $e^V$ piece is fully determined by the gauge invariance constraint and its derivative terms will not enter Hilbert series as independent fields/spurions.

Now that we've established the relevant operator building blocks (PE arguments), our next step is to determine whether we need to consider the superpotential terms ($F$-terms), K\"ahler terms ($D$-terms) or both when thinking about higher dimensional operators. As in Ref.~\cite{Delgado:2022bho}, we find we can focus almost entirely on the K\"ahler term. First, for superpotential terms with no field strengths, the same logic as Ref.~\cite{Delgado:2022bho} applies. For superpotential terms containing one or more field strength, the constraints from chirality and gauge invariance are strict, and one can readily verify that it's impossible to add $\Phi$ or $\Phi^\dagger$ charged under the gauge symmetry without violating one or both constraints.  A caveat in this reasoning is if there exists a neutral chiral superfield $\Phi_0$\footnote{In principle, this can be a product of arbitrary number of superfields, as long as it carries neutral charge and is chiral.}, which isn't forbidden from the superpotential solely by gauge invariance. In this case one can form an $F$-term as $\int d^2\theta (W_\alpha W^\alpha)\Phi_0$. However, in this case one cannot add any additional superderivatives, otherwise chirality will be violated. This point is easily realized by noticing that a chiral superfield will no longer be chiral once it carries superderivatives in front of it.  As a result, we will only focus on K\"ahler terms.

In Eq.~\eqref{eq:superPEarg}, we have been a bit sloppy with the index $n$ on the derivative powers, as derivative powers of a superfield that reduce by EOM should be omitted from the PE. This can be accomplished compactly by using the short conformal representation appropriate to the field in question: Eq.~\eqref{eq:scal} for $\Phi$,  Eq.~\eqref{eq:ferm} for $\nabla_\alpha \Phi$ and $W_\alpha$, Eq.~\eqref{eq:spin1} for $\nabla_\alpha W_\beta$, etc. The fact that we're considering superfields or gauge theories has no effect on this step. Similarly, the $P^{-1}(P,Q,\alpha,\beta,z)$ factor which removes IBP redundancies carries over from Ref.~\cite{Delgado:2022bho} exactly. IBP redundancies only deal with the derivatives (e.g. they shuffle among operators with the same quantum numbers), or, to be more precise, the Lorentz structure of derivatives. Therefore, the form is independent of whether we are working with normal superderivatives or covariant superderivatives.

With the architecture of the supersymmetric gauge theory in place, the last item we must deal with is the second subtlety mentioned earlier -- how to treat the gauge charges of fields in terms of characters. Recall, the issue is that the gauge transformation parameters are superfields rather than real valued functions. Naively proceeding as in the non-supersymmetric case leaves $e^V$ carrying no weight in the PE, meaning one can add arbitrary powers of it to any operator.  To account for the complex nature of the supersymmetric gauge parameter $\Lambda$ we label fields with {\it two} gauge charges, one corresponds to $\Lambda$ and the other corresponds to $\Lambda^\dagger$ (said another way, as the real and imaginary parts of $\Lambda$). Under this parametrization, the representation of a general superfield is given by $S\sim (j_1,j_2;g_1,g_2;z)$, where $j_1, j_2$ label the Lorentz group representation and $g_1,g_2$ represent charges under $\Lambda,\Lambda^\dagger$, respectively. For example, $\Phi_l\sim(0,0;-t_l,0),\Phi_l^\dagger\sim(0,0;0,t_l)$, etc. To project a gauge-invariant operator, we need two $U(1)$ group parameters, as well as two $U(1)$ Haar measures.

Putting the pieces together, let us apply the method to study the operator basis for an abelian, supersymmetric $U(1)$ gauge theory with a single flavor of matter superfields. For simplicity, we'll take the $U(1)$ charge to be $t_l = 1$ and $R[\Phi] = 0$. Here, the explicit form for the Hilbert series is\footnote{Here we have neglected $\Delta \mathcal H$ terms. As explained in the text, these only contribute to operators with mass dimension $d \le 4$.}:
\begin{equation}\label{abelian hs form}
\begin{split}
&\mathcal{H}(P,Q,\Phi,\Phi^\dagger,W_\alpha,\overline{W}^{\dot{\alpha}},e^V)\\
=&\int d\mu_{Lorentz}d\mu_{gauge}d\mu_{U_R(1)}P^{-1}(P,Q,\alpha,\beta,z)PE[\mathcal{I}(\Phi,\Phi^\dagger,W_\alpha,\overline{W}^{\dot{\alpha}},e^V)],
\end{split}
\end{equation}
where the Haar measures are
\begin{subequations}
\begin{align}
&d\mu_{Lorentz}=\frac{1}{(2\pi i)^2}\oint_{|\alpha|=1} \frac{d\alpha}{\alpha}(1-\alpha^2)\oint_{|\beta|=1} \frac{d\beta}{\beta}(1-\beta^2),\\
&d\mu_{gauge}=\frac{1}{(2\pi i)^2}\oint_{|g_1|=1} \frac{dg_1}{g_1}\oint_{|g_2|=1} \frac{dg_2}{g_2},\\
&d\mu_{U_R(1)}=\frac{1}{2\pi i}\oint_{|z|=1} \frac{dz}{z},
\end{align}
\end{subequations}
and 
\begin{align}
\label{ex:peargs}
& \mathcal{I}(\Phi,\Phi^\dagger,W_\alpha,\overline{W}^{\dot{\alpha}},e^V)_{\text{bos}} =\Phi g_1^{-1}\tilde{\chi}_{(0,0)} + \Phi^\dagger g_2\tilde{\chi}_{(0,0)} + \\
& ~~~~~~~~~~~~~~~~~~~~~~~~~~~~~~~~~~~~~~~~~~~~~~~~~P(DW_\alpha) \tilde{\chi}_{(1,0)}+ Q(\overline D\overline{W}^{\dot \alpha}) \tilde{\chi}_{(0,1)} + e^Vg_1g_2^{-1} \nonumber \\
& \mathcal{I}(\Phi,\Phi^\dagger,W_\alpha,\overline{W}^{\dot{\alpha}},e^V)_{\text{ferm}} = P(D\Phi) g_1^{-1}z^{-1}\tilde{\chi}_{(\frac{1}{2},0)}+ Q(\overline D\Phi^\dag) g_2z\tilde{\chi}_{(0,\frac{1}{2})} + W_\alpha z \tilde{\chi}_{(\frac{1}{2},0)} + \overline{W}^{\dot \alpha}z^{-1}\tilde{\chi}_{(0,\frac{1}{2})} \nonumber
\end{align}
are the arguments of the bosonic and fermionic plethystic exponentials\footnote{Note that, in Eq.~\eqref{ex:peargs}, the indices $_\alpha, ^{\dot\alpha}$ are purely cosmetic. The Lorentz transformation properties they imply are carried by the characters $\tilde \chi$}. The conformal characters $\tilde{\chi}$ are given in \eqref{eq:confchar} and $g_1, g_2$ are the group characters for the two $U(1)$ groups; $\Phi$ is accompanied by the group parameter for one $U(1)$, $\Phi^\dag$ is accompanied by the group parameter for the other $U(1)$, and $e^V$ appears with both group parameters. Note that, like the chiral superfields, the field strength superfields contribute one term to the bosonic PE and one term to the fermionic PE, and we have used new spurions ($(DW)$ and $(\overline{DW}) )$ to represent odd powers of superderivatives on the field strengths. 

Say we want to study the specific set of operator $\mathcal{O}(\Phi\Phi^\dagger W_\alpha^2\overline{W}^{\dot{\alpha}}\nabla_\alpha^2\overline{\nabla}_{\dot{\alpha}})$, meaning operators built from one chiral superfield $\Phi$, one antichiral superfield $\Phi^\dagger$, two $W_\alpha$'s, one $\overline{W}^{\dot{\alpha}}$ and three derivatives $\nabla_\alpha^2\overline{\nabla}_{\dot{\alpha}}$. To find/project such operators, in terms of spurions ($\mathcal{O}(\Phi\Phi^\dagger W_\alpha^2\overline{W}^{\dot{\alpha}}P^2Q)$), we simply select the terms in the expansion, i.e. $\mathcal{H}(P,Q,\Phi,\Phi^\dagger,W_\alpha,\overline{W}^{\dot{\alpha}},e^V)|_{\Phi\Phi^\dagger W_\alpha^2\overline{W}^{\dot{\alpha}}P^2Q}$. One should notice that we do not put any constraints on $e^V$ since this is totally fixed by gauge invariance. The result consists of several operators,  but if our aim is to simply count the number instead of knowing the details, we can take all spurions to be $1$ to get the number we want. In this case, we'll get $7-7+1=1$ term left in the operator basis, which indicates that only one term is independent at order $\mathcal{O}(\Phi\Phi^\dagger W_\alpha^2\overline{W}^{\dot{\alpha}}\nabla_\alpha^2\overline{\nabla}_{\dot{\alpha}})$. One can check this by using a direct, brute force calculation along the lines of Ref~\cite{Lehman:2015coa, Delgado:2022bho}, and we give the details in Appendix \ref{brute force}.

\subsection{Non-Abelian Case}\label{non abelian}

The main lesson from the supersymmetric abelian gauge theory is that the $P^{-1}$ piece carries over unchanged from theories with purely chiral/antichiral fields. Said another way, $P^{-1}$ only involves the Lorentz/derivative structure of operators and is blind to how the internal symmetry is handled. Given that non-supersymmetric Hilbert series exhibit the same behavior, this `factorization' is not surprising

With this understanding, the non-abelian case is a direct generalization, where one replaces $U(1)$ gauge symmetry with a more complicated group, $SU(2),SU(3)$, etc.. The transformations in Eq.~\eqref{u1trans} now become:
\begin{equation}\label{non abelian trans}
\Phi_l\rightarrow\Phi'_l=e^{-i\Lambda}\Phi_l;\ \Phi_l^{\dagger}\rightarrow\Phi_l^{'\dagger}=e^{i{\Lambda^\dagger}}\Phi^{\dagger}_l,
\end{equation} 
where $\Lambda=T^a\Lambda^a$ is a matrix, with $T^a$ the hermitian generators of the gauge group. As before, $\Lambda^a$ are chiral superfields such that the transformations \eqref{non abelian trans} will not change the chirality conditions.  The vector superfield $V$ becomes $V=T^aV^a$, and one can verify that the transformation law $e^V\rightarrow e^{V'}=e^{-i\Lambda^\dagger}e^Ve^{i\Lambda}$ still holds in non-abelian case. The validity of this transformation indicates that \eqref{u1 dterm} is still a candidate for the $D$-term. 

Non-abelian superfield strengths are not gauge invariant, rather they transform covariantly, i.e. $W_\alpha\rightarrow W'_\alpha=e^{-i\Lambda^\dagger}W_\alpha e^{i\Lambda}$. In addition, the covariant superderivatives \eqref{derivative} need to be modified for the non-abelian case:
\begin{subequations}\label{nonabelian derivative}
\begin{align}
&\nabla_\alpha\Phi\equiv e^{-V}D_\alpha(e^V\Phi),\ \nabla_\alpha\Phi^\dagger\equiv D_\alpha\Phi^\dagger=0,\\
&\overline{\nabla}_{\dot{\alpha}}\Phi^\dagger\equiv e^{V}\overline{D}_{\dot{\alpha}}(e^{-V}\Phi^\dagger),\ \overline{\nabla}_{\dot{\alpha}}\Phi\equiv \overline{D}_{\dot{\alpha}}\Phi=0,
\end{align}
\end{subequations}
Although the action of these covariantized derivatives are different from the ones in abelian case, their representation under Lorentz group doesn't change at all. Additionally, covariant derivative acting on $e^V$ still vanishes as in the abelian case. Therefore the form of the Hilbert series will not change and the only difference is the way one labels each superfield. 

Recall that, in the abelian case, we need two parameters to label the single $U(1)$ gauge symmetry, due to the fact that superfields transform not up to a phase, but instead in terms of superfields. As a result one needs twice the number of group parameters to label different superfields. The same argument holds in the non-abelian case, so we must indicate two different representations for each superfield, $S \sim (l_1, l_2; r_1,r_2; z)$, where $r_1, r_2$ label the representations under the two copies of the non-abelian group (and $z$ is the R-charge). Taking $SU(2)$ as an example with matter in the fundamental representation, we have $\Phi \sim (0,0; 2, 0; r)$, $\Phi^\dag \sim (0,0; 0,2; r^{-1})$. With these building blocks, we can construct the operator basis for non-abelian case follow the same steps used as in the abelian case.

As an example, let's consider the simplest non-abelian case -- gauge group $SU(2)$ -- with matter content consisting of a single chiral superfield flavor $\Phi,\Phi^\dagger$ in the fundamental representation. The Hilbert series in this case looks identical to \eqref{abelian hs form}:
\begin{align}
& \mathcal{H}(P,Q,\Phi,\Phi^\dagger,W_\alpha,\overline{W}^{\dot{\alpha}},e^V) =  \\
& ~~~~~~~~~~~~~~~~~~~~~~~~~  \int d\mu_{Lorentz}d\mu_{gauge}d\mu_{U_R(1)}P^{-1}(P,Q,\alpha,\beta,z)PE[\mathcal{I}(\Phi,\Phi^\dagger,W_\alpha,\overline{W}^{\dot{\alpha}},e^V)], \nonumber 
\end{align}
but the Haar measures are different:
\begin{equation}
d\mu_{gauge}=\frac{1}{(2\pi i)^2}\oint_{|g_1|=1} \frac{dg_1}{g_1} (1-g_1^2) \oint_{|g_2|=1} \frac{dg_2}{g_2}(1-g_2^2),
\end{equation}
as we need two $SU(2)$ measures (with group parameters $g_1, g_2$) instead of two $U(1)$ measures. The argument of the PE, again for the choice $R[\Phi] = 0$, is: 
\begin{align}
\mathcal{I}(\Phi,\Phi^\dagger,W_\alpha,\overline{W}^{\dot{\alpha}},e^V)_{\text{bos}}&=\Phi (g_1+\frac{1}{g_1})\tilde{\chi}_{(0,0)}+\Phi^\dagger (g_2+\frac{1}{g_2})\tilde{\chi}_{(0,0)}  +  \\
& ~~~~~~~~~~~~~~~~~~~~~~~~~~~P(DW_\alpha)\tilde{\chi}_{(1,0)} + Q\overline{DW}^{\dot \alpha}\tilde{\chi}_{(0,1)} +e^V(g_1+\frac{1}{g_1})(g_2+\frac{1}{g_2}) \nonumber \\
\mathcal{I}(\Phi,\Phi^\dagger,W_\alpha,\overline{W}^{\dot{\alpha}},e^V)_{\text{ferm}}&= P(D\Phi) (g_1+\frac{1}{g_1})z^{-1}\tilde{\chi}_{(\frac{1}{2},0)}+Q(\overline D\Phi^{\dag}) (g_2+\frac{1}{g_2})z\tilde{\chi}_{(0,\frac{1}{2})}+ \nonumber \\
&~~~~~~~~~~~~~~~~~~~~~~~~~~~~~~~~~~~~~~~~~~~~~~~~~W_\alpha z \tilde{\chi}_{(\frac{1}{2},0)} + \overline{W}^{\dot \alpha}z^{-1}\tilde{\chi}_{(0,\frac{1}{2})}, \nonumber
\end{align}
with conformal characters $\tilde{\chi}$ given in \eqref{eq:confchar}.

\section{Conclusion and Discussion}\label{conclusion}
In this paper we show how to count operators in $N=1$ supersymmetric gauge theories. We provide two examples: the abelian case and the non-abelian case with the explicit and detailed Hilbert series constructions. The main difference between supersymmetric gauge case and the non-supersymmetric gauge theory is that one has to double the number of gauge group parameters in order to give the correct $e^V$ structure. Although the two examples given in the text only deal with single flavor in the fundamental representation, the approach can be extended to include more flavors of any representation and with arbitrary number of vector/chiral/antichiral superfields.  A shortcoming of the Hilbert series method (for both supersymmetric and non-supersymmetric theories), however, is that it does not explicitly give the detailed structure of the operators -- meaning where derivatives are placed and how indices are contracted -- so additional work is required if one wants the actual operator forms. Nevertheless, being able to count the operators in supersymmetric gauge theories will benefit future EFT study of supersymmetric theories. 

In non-supersymmetric theories, one can find the explicit form of operators using the Young Tableau construction \cite{Henning:2019enq, Henning:2019mcv, Ma:2019gtx, Durieux:2019siw, Dong:2021yak,Li:2022tec,Fonseca:2019yya,Li:2020gnx}, as mentioned earlier. Progress on a generalization of the Young Tableau technique to supersymmetric theories will appear in future work~\cite{susy_ssyt}.

\acknowledgments

This is partially supported by the National Science Foundation under Grant Number PHY-2112540.

\appendix

\section{Cross-check of counting for $\mathcal{O}(\Phi\Phi^\dagger W_\alpha^2\overline{W}^{\dot{\alpha}}\nabla_\alpha^2\overline{\nabla}_{\dot{\alpha}})$ operators}\label{brute force}

There are seven operators at $\mathcal{O}(\Phi\Phi^\dagger W_\alpha^2\overline{W}^{\dot{\alpha}}\nabla_\alpha^2\overline{\nabla}_{\dot{\alpha}})$:
\begin{align}
& a_1 \equiv\nabla_\alpha\Phi e^V\nabla^\alpha W_\beta W^\beta\overline{\nabla}_{\dot{\alpha}}\Phi^\dagger\overline{W}^{\dot{\alpha}},\quad  a_2  \equiv\Phi e^V\nabla^\alpha W_\beta\nabla_\alpha W^\beta\overline{\nabla}_{\dot{\alpha}}\Phi^\dagger\overline{W}^{\dot{\alpha}}, \nonumber \\
& a_3 \equiv\nabla^\alpha\Phi e^V W_\beta W^\beta\nabla_{\alpha\dot{\alpha}}\Phi^\dagger\overline{W}^{\dot{\alpha}},\quad a_4 \equiv\Phi e^V\nabla^\alpha W_\beta W^\beta\nabla_{\alpha\dot{\alpha}}\Phi^\dagger\overline{W}^{\dot{\alpha}}, \\ 
& a_5 \equiv\nabla^\alpha\Phi e^V\nabla_{\dot{\alpha}\alpha} W_\beta W^\beta\Phi^\dagger\overline{W}^{\dot{\alpha}},\quad  a_6  \equiv\nabla_{\dot{\alpha}\alpha}\Phi e^V\nabla^\alpha W_\beta W^\beta\Phi^\dagger\overline{W}^{\dot{\alpha}}, \nonumber \\
& ~~~~~~~~~~~ a_7  \equiv\Phi e^V\nabla_{\dot{\alpha}\alpha} W_\beta\nabla^\alpha W^\beta\Phi^\dagger\overline{W}^{\dot{\alpha}}. \nonumber 
\end{align}
There are five IBP relations among these seven operators from the $\nabla_\alpha$ branch:
\begin{align}
&c_1\equiv\nabla^\alpha(\nabla_\alpha\Phi e^V W_\beta W^\beta\overline{\nabla}_{\dot{\alpha}}\Phi^\dagger\overline{W}^{\dot{\alpha}})\sim-2a_1+a_3\sim0 \nonumber \\
&c_2\equiv\nabla^\alpha(\Phi e^V\nabla_\alpha W_\beta W^\beta\overline{\nabla}_{\dot{\alpha}}\Phi^\dagger\overline{W}^{\dot{\alpha}})\sim-a_1-a_2+a_4\sim0 \nonumber\\
&c_3\equiv\nabla^\alpha(\Phi e^V W_\beta W^\beta\nabla_{\alpha\dot{\alpha}}\Phi^\dagger\overline{W}^{\dot{\alpha}})\sim a_3+2a_4\sim0\\
&c_4\equiv\nabla^\alpha(\nabla_{\dot{\alpha}\alpha}\Phi e^V W_\beta W^\beta\Phi^\dagger\overline{W}^{\dot{\alpha}})\sim 2a_6\sim0 \nonumber\\
&c_5\equiv\nabla^\alpha(\Phi e^V \nabla_{\dot{\alpha}\alpha} W_\beta W^\beta\Phi^\dagger\overline{W}^{\dot{\alpha}})\sim a_5-a_7\sim0, \nonumber
\end{align}
and two IBP relations from the $\overline{\nabla}_{\dot{\alpha}}$ branch:
\begin{align}
&c_6\equiv\overline{\nabla}_{\dot{\alpha}}(\nabla_\alpha\Phi e^V\nabla^\alpha W_\beta W^\beta\Phi^\dagger\overline{W}^{\dot{\alpha}})\sim -a_6-a_5-a_1\sim0 \nonumber \\
&c_7\equiv\overline{\nabla}_{\dot{\alpha}}(\Phi e^V\nabla_\alpha W_\beta\nabla^\alpha W^\beta\Phi^\dagger\overline{W}^{\dot{\alpha}})\sim 2a_7-a2\sim0
\end{align}
The seven $c_i$ are not all independent, as $c_1+2c_2=c_3+c_4+2c_6+2c_5+c_7$. Taking this into account, there are only six relations among the seven operators, leaving one independent $a_i$ term.

\section{Characters and Haar Measures}
\label{app:characters}

Here we list a few characters in group $U(1),\ SU(2)$ and $SU(3)$ that are used in this paper. And we also list the related Haar measure for each group. For further characters, please refer to \cite{Hanany:2008sb}.

The characters are given by:
\begin{subequations}
\begin{align}
&\chi_{U(1)}=e^Q,\\
&\chi_{SU(2) fund}=z+\frac{1}{z},\\
&\chi_{SU(3) fund}=z_1+\frac{z_2}{z_1}+\frac{1}{z_2},
\end{align}
\end{subequations}
where $Q,\ z,\ z_1,\ z_2$ are related group parameters.

The Haar measures are given by: 
\begin{subequations}
\begin{align}
&\int d\mu_{U(1)}=\frac{1}{2\pi i}\oint_{|z|=1} \frac{dz}{z},\\
&\int d\mu_{SU(2)}=\frac{1}{2\pi i}\oint_{|z|=1} \frac{dz}{z}(1-z^2),\\
&\int d\mu_{SU(3)}=\frac{1}{(2\pi i)^2}\oint_{|z_1|=1} \frac{dz_1}{z_1}\oint_{|z_2|=1} \frac{dz_2}{z_2}(1-z_1z_2)(1-\frac{z_1^2}{z_2})(1-\frac{z_2^2}{z_1}).
\end{align}
\end{subequations}

\bibliographystyle{utphys}
\bibliography{ref}

\end{document}